\documentclass[twocolumn,showpacs,preprintnumbers,amsmath,amssymb,a4paper,superscriptaddress]{revtex4}

\usepackage{graphicx}
\usepackage{dcolumn}
\usepackage{bm}
\usepackage{color}

\begin{document}

\title{Resonant d-wave scattering in harmonic waveguides}

\author{P. Giannakeas}
\email{pgiannak@physnet.uni-hamburg.de}
\affiliation{Zentrum f\"{u}r Optische Quantentechnologien, Universit\"{a}t Hamburg, Luruper Chaussee 149, 22761 Hamburg,
Germany,}

\author{V.S. Melezhik}
\email{melezhik@theor.jinr.ru}
\affiliation{Bogoliubov Laboratory of Theoretical Physics, Joint Institute for Nuclear Research,
Dubna, Moscow Region 141980, Russian Federation,}

\author{P. Schmelcher}
\email{pschmelc@physnet.uni-hamburg.de}
\affiliation{Zentrum f\"{u}r Optische Quantentechnologien, Universit\"{a}t Hamburg, Luruper Chaussee 149, 22761 Hamburg,
Germany,}

\date{\today}

\begin{abstract}
We observe and analyze d-wave resonant scattering of bosons in tightly confining harmonic waveguides.
It is shown that the d-wave resonance emerges in the quasi-1D regime as an imprint of a 3D d-wave shape resonance.
A scaling relation for the position of the d-wave resonance is provided.
By changing the trap frequency, ultracold scattering can be continuously tuned from s-wave to d-wave resonant behavior.
The effect can be utilized for the realization of ultracold atomic gases interacting via higher partial waves and opens a novel possibility for studying strongly correlated atomic systems beyond s-wave physics.
\end{abstract}


\maketitle
\section{Introduction}

The control of the interaction via resonances in atomic and molecular many-body systems is of major importance in the field of ultracold quantum gases.
Magnetic and optical Feshbach resonances are well established tools of experimentally tuning the interactions of atomic ensembles \cite{inouye98,theis04,tiesinga10}.
In recent years, however, an alternative way of controlling the inter-particle correlations has attracted increasing interest: the confinement-induced resonance (CIR) \cite{olshanii98,haller09,gunter05}.
It emerges due to the interplay of atom-atom scattering with the tightly confining geometry of the trap, and yields drastic changes of the scattering process and its properties \cite{olshanii98,olshanii03}.
CIRs have been extensively studied e.g. in the context of three-body \cite{mora04, mora05} and four-body \cite{mora05b} scattering in a confining potential, of p-wave scattering of fermions \cite{blume04},
of scattering in mixed dimensions \cite{nishida10} or multichannel scattering in cylindrical confinement \cite{saeidian08}.
For distinguishable atoms, two major effects are the so-called dual CIR yielding a complete suppression of quantum scattering, and resonant molecule formation in tight waveguides \cite{kim06,melezhik09}.
Remarkable experimental progress has lead to the observation of CIRs for both bosons \cite{kinoshita04,paredes04,haller09} and fermions \cite{gunter05}, relying on the control of s-wave and p-wave interactions in bosonic and fermionic low-dimensional systems, respectively.
It is an intriguing perspective to extend the resonant scattering physics to higher partial wave interactions in quantum gases, since the latter is expected to provide novel many-body phenomena.
A promising example is the possibility that higher partial wave interactions are responsible for high-temperature superconductivity and superfluidity \cite{rey09,deb09,hofstetter02}.

In the present work we suggest a novel mechanism for resonant higher partial-wave interatomic interactions in bosonic quantum gases, based on the effect of resonant d-wave interactions for ultracold collisions in tightly confining waveguides.
We provide a resonance condition implying the dependence of the position of the d-wave resonance (DWR) on the oscillator length $a_{\perp}=\sqrt{\hbar/(\mu\omega_{\perp})}$ ($\omega_{\perp}$ is the confinement frequency) of the harmonic waveguide.
It is shown that this condition does not depend on the short-range part of the interatomic interaction but on the long-range part via the dispersion coefficient $C_{6}$.
Additionally, the strength of the d-wave interactions can be controlled by changing the width $a_{\perp}$ of the waveguide, as it is the case for s-wave scattering.
Due to the fact that DWRs occur in the background of s-wave CIRs, an adjustment of the trap width can change entirely the nature of the underlying interactions of the colliding atoms from s- to d-wave and vice versa.
We find that the DWR observed here is based on a shape resonance formed in the presence of both the centrifugal barrier and the confining trap.

\section{Hamiltonian, methodology \quad\quad\quad and setup}

We consider ultracold collisions of identical bosons in a harmonic waveguide with a transverse potential $\frac{1}{2} \mu\omega_{\perp}^2 \rho^2$ ($\rho = r \sin \theta$).
The latter permits a separation of the center of mass and relative motion yielding the following Hamiltonian for the relative motion:
\begin{equation}
H(r,\theta,\phi) = -\frac{\hbar^2}{2\mu}\frac{\partial ^{2}}{\partial r^{2}}+\frac{\hbar^2}{2\mu}
\frac{L^{2}(\theta,\phi)}{r^{2}}+\frac{1}{2}\mu\omega_{\perp}^2\rho^2+V(r)
\label{1}
\end{equation}
$V(r)$ is the interatomic potential, where $r$ is the relative radial coordinate and $\mu = m/2$ is the reduced mass of the two bosons.
The boundary conditions for quasi-1D scattering in the waveguide read, for $z=r\cos\theta\rightarrow \pm\infty$,
\begin{equation}
\psi(r,\theta)\rightarrow\left[\cos(kr\cos\theta)+f_{0}e^{ikr\mid\cos\theta\mid}\right]\Phi_{0,0}(r\sin\theta)
\label{2}
\end{equation}
where $f_{0}$ is the elastic scattering amplitude of the ground transversal channel, $\Phi_{0,0}(r\sin\theta)$ is the ground-state wave-function of the 2D harmonic oscillator with transversal energy $\varepsilon_{\perp}=\hbar\omega_{\perp}$, $k=\sqrt{2\mu(\varepsilon-\hbar\omega_{\perp})}$ is the relative momentum of the colliding pair, and $\varepsilon=\varepsilon_{\parallel}+\varepsilon_{\perp}$ is the energy of the relative two-body motion written as a sum of the longitudinal $\varepsilon_{\parallel}$ and transversal $\varepsilon_{\perp}$ energies.
We focus on the single mode regime for which the energy lies below the first excited transversal energy level, $\hbar\omega_{\perp} \leq \varepsilon < 3\hbar\omega_{\perp}$.
Obviously, the scattering state (\ref{2}) is parity $(\bf{r}\rightarrow -\bf{r})$ symmetric, corresponding to the case of two colliding identical bosons.

The Hamiltonian allows for a separation of the azimuthal motion, corresponding to the conservation of the projection of the angular momentum onto the symmetry axis of the waveguide.
Consequently, we obtain a 2D scattering problem [Eqs.~(\ref{1}) and(\ref{2})] in spherical coordinates, which is solved by employing the discrete-variable method suggested in Refs.~\cite{melezhik91,saeidian08}, with the radial part of the Schr\"odinger equation discretized using a B-spline basis \cite{carl78,lee82}.
The interatomic interaction $V(r)$ is modeled via the Lennard-Jones 6-12 potential $V(r)=C_{12}/r^{12}-C_{6}/r^{6}$.
We hereafter employ the units $m_{\text{Cs}}/2 = \hbar = \omega_0 = 1$, where $m_{\text{Cs}}$ is the mass of the Cs atom and $\omega_0 = 2\pi\times 10$ MHz in SI units.
The following investigations are performed for the dispersion coefficients $C_{6}$ defining the long-range part of the interatomic potential for three different atomic species, Cs, Rb and Sr \cite{derevianko01,derevianko02}.
The longitudinal energy is set to $\varepsilon_{\parallel} = 2 \times 10^{-6}$ and the transversal energy is varied within the interval $2 \times 10^{-4} \leq \varepsilon_{\perp} \leq 2 \times 10^{-2}$, corresponding to a range $4\pi~\text{KHz} \leq \omega_{\perp} \leq 400\pi~\text{KHz}$ (SI) for the waveguide confinement frequency.
We thereby focus on the low energy regime, characterized by $ka_{\perp} \ll 1$.

\begin{figure}[!t]
\includegraphics[width=0.7\columnwidth]{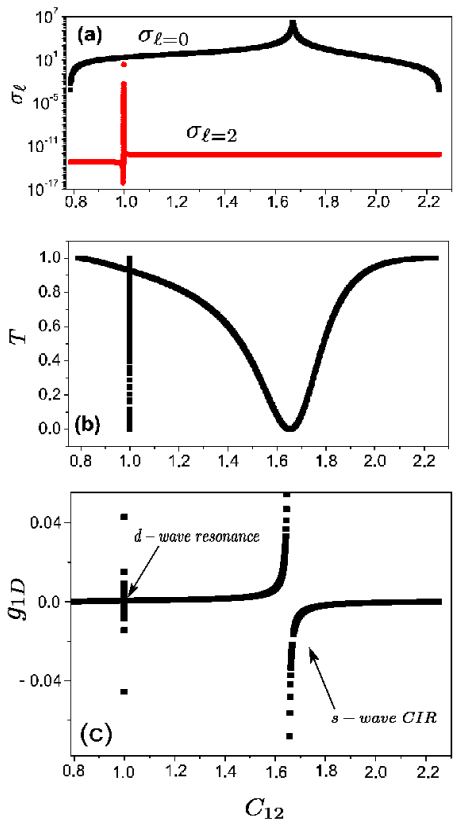}
\caption{(color online) (a) Partial cross sections for s- and d-wave scattering in free space, (b) transmission coefficient $T$ in the waveguide and (c) interaction strength $g_{1D}$ as a function of the parameter $C_{12}$ ($C_{6}=28.82$ for Cs).}
\label{fig1}
\end{figure}

Important quantities of our analysis are the scattering amplitude $f_{0}$ and the transmission coefficient
\begin{equation}
T=\lvert1+f_0\rvert^2
\label{3}
\end{equation}
which we analyze in the limit of very small longitudinal collision energies for which the s-wave CIR was initially defined \cite{olshanii98}.

\section{Results, analysis and discussion}

Let us first briefly address free space Cs-Cs collisions ($C_{6}=28.82$ \cite{derevianko01,derevianko02}) in the energy range for which s-wave scattering dominates the elastic scattering process.
By varying $C_{12}$ (which changes the short range part of the interatomic interaction) we observe s- and d-wave resonances in the corresponding partial cross sections [see Fig.~\ref{fig1}(a)].
As expected, the width of the d-wave resonance is by orders of magnitude smaller than that of the s-wave resonance.
In the presence of the waveguide [see Fig.~\ref{fig1}(b)] the free near resonant s-wave scattering turns into a s-wave confinement-induced resonance \cite{olshanii98,olshanii03,saeidian08,peano05,kim06,melezhik07}, which appears as a broad dip in the transmission coefficient with a zero-valued minimum at $C_{12} \cong 1.65$, coinciding with a divergence of the interaction strength $g_{1D} = \lim_{\varepsilon_{\parallel}\rightarrow 0} ({k\text{Re}f_{0}}/{\text{Im}f_{0}})$.
In order to compare the numerically calculated position of the s-wave CIR with the analytical result \cite{olshanii98} we map the $C_{12}$ parameter onto the s-wave scattering length $a_{s}$ in free space and obtain the ratio $a_{s}(C_{12} \cong 1.65) / a_{\perp} \cong 0.65$ which is in good agreement with the analytical prediction $a_{s}/a_{\perp} \cong 0.68$.

The d-wave resonance persists also in the waveguide, where it appears as a strongly varying transmission coefficient between the limiting values 0 and 1 on top of the s-wave scattering background [Fig.~\ref{fig1}(b)].
As in the case of the s-wave CIR, we observe a divergence of $g_{1D}$ at the resonant value $C_{12}=0.9988731$ [Fig.~\ref{fig1}(c)].

\begin{figure}[!t]
\includegraphics[width=0.9\columnwidth]{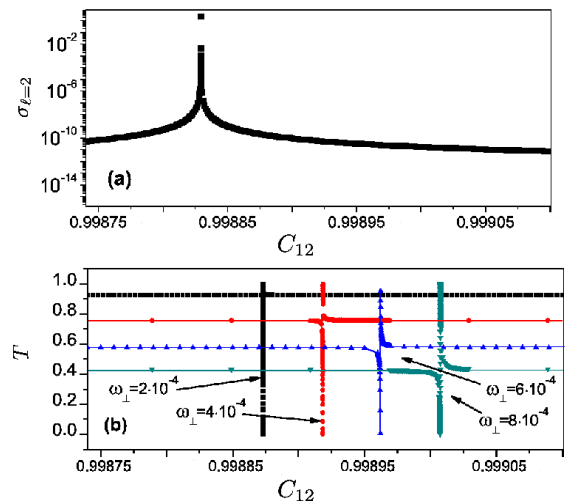}
\caption{(color online) (a) Partial cross section $\sigma_{\ell = 2}$ for d-wave scattering in free space and (b) transmission coefficient for several values of the confinement frequency $\omega_{\perp}$, as a function of the parameter $C_{12}$ around the d-wave resonance.}
\label{fig2}
\end{figure}

The transversal confinement leads to a shift of the d-wave resonance when passing from free space to confined scattering, as is clearly seen in the high resolution graph of Fig.~\ref{fig2}.
This shift, as well as the corresponding width, increase with the trap frequency $\omega_{\perp}$ [see Fig.~\ref{fig2}(b)].
We also observe a strong suppression of the s-wave background with increasing $\omega_{\perp}$.
The above behavior of the transmission coefficient $T(\omega_{\perp}$,$C_{12})$ under the action of the confining potential can be interpreted in terms of a strong coupling of s- and d-waves in harmonic traps, since the trap potential can be represented as a sum of the Legendre polynomials:
\begin{equation}
 \frac{1}{2}\mu\omega_{\perp}^2\rho^2=\frac{1}{6}\mu\omega_{\perp}^2r^2[2P_0(\cos\theta)-P_2(\cos\theta)]
\label{4}
\end{equation}
Consequently, the confining potential induces a coupling between partial waves $\ell$ and $\ell + 2$ in the course of the atomic collisions in the trap.
This coupling can be tuned by changing the trap frequency.

\begin{figure}[!t]
\includegraphics[width=0.6\columnwidth]{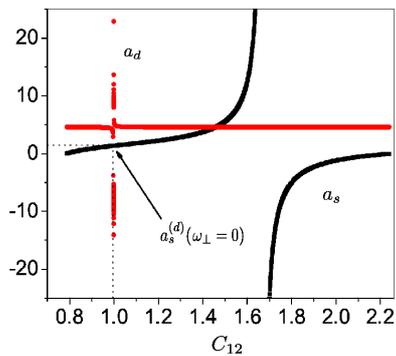}
\vspace{-.3cm}
\caption{(color online) s- and d-wave scattering lengths $a_{s}$ and $a_{d}$ in free space as a function of the $C_{12}$ parameter, where $a_d$ is defined through $a_{\ell}^{2\ell + 1} = -\tan\delta_{\ell} / k^{2\ell + 1}$ \cite{stock05} for $\ell = 2$. $a_{s}^{(d)}(\omega_{\perp}=0) \equiv a_{s,0}^{(d)}$ is the value of the s-wave scattering length at the point of divergence of the d-wave scattering length $a_{d}$.}
\label{fig3}
\end{figure}

We now analyze more precisely the condition for the occurrence of the DWR.
The experimental determination of the d-wave scattering length and in particular of the position of the corresponding resonance is a difficult task.
However, in the following we will show the existence of a very useful scaling relation for the position of the DWR that provides us with an accurate estimate of its position.

\begin{figure}[!t]
      \includegraphics[width=0.7\columnwidth]{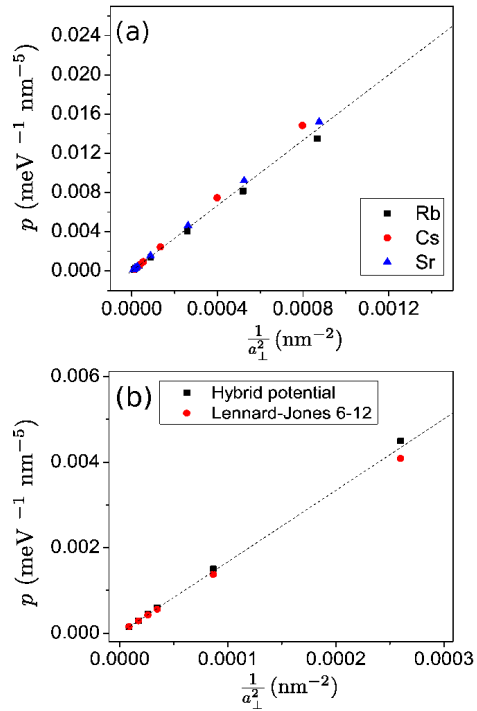}
\caption{ \label{fig4} (color online) Quotient $p = \Delta a_s^{(d)} / C_6$ as a function of $1/a_{\perp}^{2}$ (a) for different values of $C_6$ corresponding to the atomic species Cs, Rb and Sr and (b) for Rb atoms with different interatomic potentials (see text), for confinement frequencies $4\pi \text{KHz} \leq \omega_{\perp} \leq 400\pi \text{KHz}$. Dashed lines correspond to the resonance condition given in
Eq.~(\ref{5}).}
\end{figure}

Let us firstly introduce the quantity $a^{(d)}_{s}$, which is the free space s-wave scattering length $a_{s}$ at the position of the free space d-wave shape resonance.
This is well defined since $a_{s}$ in general varies smoothly in the vicinity of the d-wave resonance (see Fig.~\ref{fig3}).
Similarly in \cite{londono10} the crucial role of $a_s$
for the analysis of the ultracold scattering near the shape resonances in
free space  was shown. We denote its value $a^{(d)}_{s}(\omega_{\perp}=0)$ in free space by $a^{(d)}_{s,0}$.
The next step is to determine the shift $\Delta a_s^{(d)} = a_s^{(d)} - a_{s,0}^{(d)}$ of the position of the DWR in the waveguide $a_s^{(d)} = a_s^{(d)}(\omega_{\perp}\neq 0)$ with respect to the resonance position in free space, in the limit $\varepsilon_{\parallel}\rightarrow 0$.
Note that $a_{s}$ is altered here by varying the $C_{12}$ parameter, while experimentally it may be achieved by employing e.g. magnetic Feshbach resonances \cite{haller09,tiesinga10}.
In Fig.~\ref{fig4}(a) we show the dependence of the quotient $p = \Delta a_s^{(d)} / C_6$ on $1 / a_{\perp}^2$ for the three different atomic species Cs, Rb and Sr with corresponding $C_6$ coefficients \cite{derevianko01,derevianko02}.
By fitting the numerical data we find this dependence to be rather accurately described by the linear relation
\begin{equation}
p  ~ = ~ \frac{\Delta a_s^{(d)}}{C_{6}} ~ = ~ 16.7~\frac{1}{a_{\perp}^2}~(\text{meV}^{-1}\text{nm}^{-5})
\label{5}
\end{equation}
for a comparatively broad range of values of $\omega_{\perp}$ for all considered $C_6$ coefficients.
This shows that the quotient $p$ changes with the confinement strength independently of the type of colliding atomic species.
In contrast to the s-wave CIR, whose position depends exclusively on the ratio $a_{s} / a_{\perp}$, the shift $\Delta a_s^{(d)}$ of the DWR depends additionally on the dispersion coefficient $C_{6}$.
This is because the Van-der-Waals potential tail determines the width of the centrifugal barrier (absent in s-wave scattering), which in turn strongly influences the DWR state.
Note that $p$ possesses the correct asymptotic value $p(a_{\perp} \rightarrow \infty) = 0$ in the limiting case of scattering in free space.
Additionally, a fitting of experimental data with the Eq.~(\ref{5})
can provide an alternative way for measuring the dispersion coefficient
$C_6$.

In order to assure that Eq.~(\ref{5}) is not specific to the model used for the interatomic interaction, we have further studied a hybrid potential which possesses a short-range part different from the Lennard-Jones 6-12 potential, reading:
\begin{equation}
V_{h}(r)=Ae^{-qr^2}-f_{c}(r)\frac{C_{6}}{r^{6}}
\label{6}
\end{equation}
where $A$ and $q$ are constants and $f_{c}(r)$ is a cut-off function \cite{marinescu94}
\begin{equation}
f_{c}(r)=\theta(r-r_{c})+\theta(r_{c}-r)\exp \left[-\left(\dfrac{r_{c}-r}{r}\right)^2\right]
\label{7}
\end{equation}
with $\theta(x)$ being the Heaviside step function and $r_{c}$ the cut-off radius.
We choose $A=5000$ and $r_{c}=1.2$ and vary the parameter $q$.
Fig.~\ref{fig4}(b) shows $p$ for Rb-Rb collisions as a function of $1 / a_{\perp}^2$ for the Lennard-Jones and the hybrid potential.
The values for $p(1 / a_{\perp}^2)$ are for both potentials in very good agreement with the scaling relation Eq.~(\ref{5}).

\begin{figure}[t!]
 \begin{center}
   \includegraphics[width=.92\columnwidth]{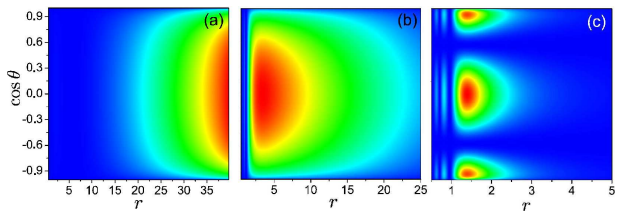}
\caption{(color online) Probability density $\lvert\Psi(r,\theta)\rvert^{2}$ for Cs-Cs collisions (a) for the off-resonant case, (b) at the s-wave CIR and (c) at the DWR, for $\omega_{\perp} = 2 \times 10^{-4}$.}
 \label{fig5}
 \end{center}
\end{figure}

Let us now inspect the probability density $\lvert\Psi(r,\theta)\rvert^{2}$ for different regimes of Cs-Cs collisions, plotted in Fig.~\ref{fig5}.
The non-resonant case [Fig.~\ref{fig5}(a)] leads to a probability density which is substantial only far from the origin $r=0$.
Fig.~\ref{fig5}(b) and (c) refer to the s-wave CIR and DWR in the waveguide, respectively.
For the s-wave CIR we observe a localization of the probability density near the origin, while the angular distribution possess no nodes.
In contrast, the DWR exhibits, as expected, two nodes in the $\theta$ coordinate.

\begin{figure}[!b]
\begin{center}
\includegraphics[width=0.6\columnwidth]{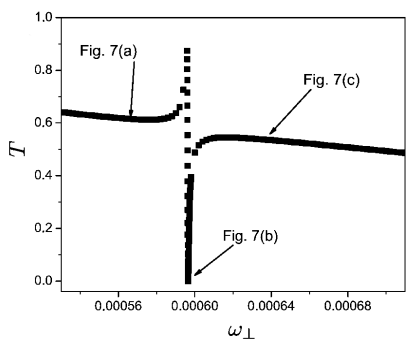}
\end{center}
\vspace{-.5cm}
\caption{(color online) Transmission coefficient for $C_{12}=0.9989621$ as a function of the confinement frequency $\omega_{\perp}$ near the DWR.}
\label{fig6}
\end{figure}

\begin{figure}[!t]
    \begin{center}
      \includegraphics[width=0.92\columnwidth]{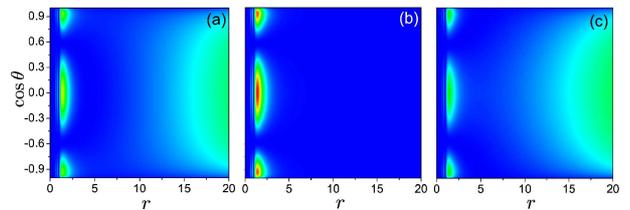}
    \end{center}
  \vspace{-.5cm}
 \caption{(color online) Probability density  $\lvert\Psi(r,\theta)\rvert^{2}$ for Cs-Cs collisions (a) $\omega=5.7\times10^{-4}$ ('left side' of the DWR),
 (b) $\omega=5.96\times10^{-4}$ (position of the DWR) and (c) $\omega=6.36\times10^{-4}$('right side' of the DWR).}
\label{fig7}
\end{figure}

Fig.~\ref{fig6} together with Fig.~\ref{fig7} illustrate how the d-wave interactions in harmonic waveguides can be controlled by altering the trap width.
Varying $\omega_{\perp}$ ($a_{\perp}$) changes the transmission coefficient $T$ near the DWR and accordingly, and so the strength of the d-wave interaction between the bosons can be tuned.
In order to further illustrate this we show in Fig.~\ref{fig7}(a)-(c) the probability density distributions for the resonant value of $\omega_{\perp}$ as well as far from the resonant region.
A weak anisotropy is observed far from resonance whereas strong anisotropy characterizes the density profile close to resonance [Fig.~\ref{fig7}(b)].
We emphasize the principal difference in the resonant behavior of the transmission coefficient near the s-wave CIR and DWR [see Figs.~\ref{fig1}(b) and ~\ref{fig6}].
Near the s-wave CIR the contribution of d-wave scattering is negligible with respect to the total 1D scattering amplitude.
However, near the DWR the d- and s-wave scattering states are strongly coupled and their interference lead to a sharp variation in the transmission coefficient between unity and zero, characterized by the Fano asymmetric lineshape \cite{fano61}.
Changing the confinement length $a_{\perp}$ one can therefore completely alter the nature of the bosonic interactions from s- to d-wave character and vice versa, since the s- and d-wave scattering coexist in the transmission coefficient.
From Fig.~\ref{fig7} one can conclude that the region of significant d-wave scattering
is rather broad with respect to the variation of $\omega_{\perp}$ and exceeds the width of the d-wave resonance in Fig.~\ref{fig6}.

\begin{figure}[!b]
\includegraphics[width=0.6\columnwidth]{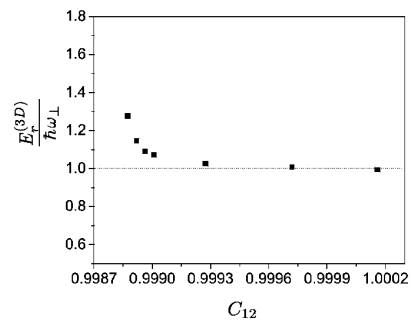}
\caption{\label{fig8} (color online) Ratio of the energy $E_r^{\text{(3D)}}$ of the resonant state in free space and the confining frequency $\omega_{\perp}$ as a function of the $C_{12}$ parameter. $\omega_{\perp}$ is varied in the range $4\pi~\text{KHz} \leq \omega_{\perp} \leq 400\pi~\text{KHz}$.}
\end{figure}

\begin{figure}[t!]
        \includegraphics[width=0.9\columnwidth]{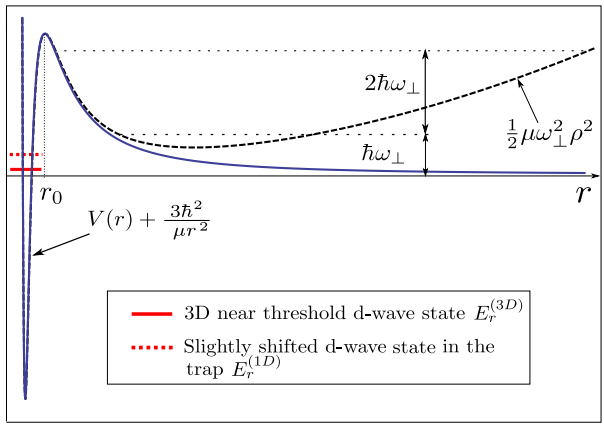}
\caption{\label{fig9} (color online) A schematic illustration of the scattering potential in free space (blue) and in the presence of confinement (black dashed).}
\end{figure}

To clarify the physical reason for the appearance of the DWR let us analyze the nature of the free space and confinement-resonant states.
Fig.~\ref{fig2}(b) provides us with the values of the $C_{12}$ parameter for each $\omega_{\perp}$ at resonance, which we then use in the 3D scattering problem to extract the corresponding
free space resonant energies $E_r^{\text{(3D)}}$.
It is found that these coincide with the the ground state energy in the harmonic confinement with increasing $C_{12}$: $E_r^{\text{(3D)}}=\hbar\omega_{\perp}$ (see Fig.~\ref{fig8}).
This allows us to the conclude upon the stability of the position of the near-threshold d-wave resonance $E_r^{\text{(3D)}}$ while adding the trap potential (see Fig.~\ref{fig9} for the following discussion).
Behind the centrifugal barrier ($r < r_0$) the trapping potential only slightly shifts the interatomic spectrum $E_r^{\text{(3D)}} \rightarrow E_r^{\text{(1D)}} \approx E_r^{\text{(3D)}}$ for $r_{0} \ll a_{\perp}$.
However, in front of the centrifugal barrier ($r > r_0$), where $V(r) \rightarrow 0$, the trap potential becomes  dominant and leads to the quantization of the interatomic continuum.

The near-threshold d-wave resonance $E_r^{\text{(3D)}} \simeq 0$ in free space ($\omega_{\perp}=0$) becomes a weakly-bound state $E_r^{\text{(1D)}} \simeq 0$ with binding energy $|E_r^{\text{(1D)}} - \hbar\omega_{\perp}| \simeq \hbar\omega_{\perp}$ referring as a threshold to the first vibrational state in the waveguide.
Thus, it becomes clear that for transforming this weakly-bound state into the DWR it has to be shifted in order to lead to an energetically degeneracy with the threshold of the transversal ground state, yielding then $E_r^{\text{(1D)}} \simeq \hbar\omega_{\perp}$.
The latter is here achieved [see Fig.~\ref{fig2}(b)] by altering the $C_{12}$ coefficient.
Experimentally, it can be realized by the magnetic Feshbach resonance technique.
Further, one can control the appearance of the DWR exclusively by tuning the trap frequency $\omega_{\perp}$ to arrive at the resonance condition $E_r^{\text{(3D)}} \simeq E_r^{\text{(1D)}} = \hbar\omega_{\perp}$ without altering the interatomic interaction $V(r)$.
The above analysis also explains the increase of the width of the DWR with increasing $C_{12}$ [see Fig.~\ref{fig2}(b)] due to a corresponding decrease of the relevant width of the centrifugal barrier.
Finally, it demonstrates that, in contrast to the s-wave Feshbach CIR \cite{note} which arises from a bound state of the first excited transversal channel, the observed DWR is a shape resonance.

\section{Brief summary}

We have demonstrated how strong d-wave interactions can develop in a harmonic waveguide and analyzed their properties in detail.
The DWR observed here is a shape-resonance as opposed to the Feshbach resonance of the s-wave CIR.
It is controllable by adjusting the frequency of the trap.
The confinement plays here a two-fold role: It can control the strength of the interactions among the bosons as well as alter the nature of the interactions from s- to d-wave character.
The DWR  effect can be analyzed and exploited in experiments analogous to \cite{haller10} with the help of the scaling relation (\ref{5}).
The observation and application of it in anisotropic waveguides represents an intriguing perspective, since in the limit of the quasi-2D regime the anisotropic profile of the d-wave interactions can be exploited to introduce novel properties into quantum gases, or more specifically speaking, strongly correlated bosonic systems.
Unconventional superfluidity and superconductivity in the presence of higher partial wave interactions are examples supposed to bear important peculiarities.

\begin{acknowledgements}
We are grateful to Dr. Elmar Haller for fruitful discussions.
V.S.M. acknowledges financial support by the Deutsche Forschungsgemeinschaft and the Heisenberg-Landau Program. P.S. thanks the Deutsche Forschungsgemeinschaft for financial support.
\end{acknowledgements}


\begin{thebibliography}{99}
\bibitem{note} We note that d-wave resonances of similar type, arising from higher channel bound d-wave states lying in the continuum of the open channel, are also present in this system; however, due to the centrifugal barrier their coupling to the continuum is very weak, which makes the detection of such d-wave Feshbach CIRs a highly demanding task.
\end{thebibliography}
\end{document}